\documentclass[12pt]{article}
\pdfoutput=1
 \usepackage[nottoc]{tocbibind}
\usepackage[colorlinks=true,linkcolor=blue,urlcolor=magenta, citecolor=blue]{hyperref}
\usepackage{fullpage}
\usepackage{amssymb,graphicx}
\usepackage{amsmath}
\usepackage[margin=0.5cm]{caption}
\usepackage{hyperref} 
\usepackage{cite}
\usepackage{upgreek}

\def\d{\partial}
\def\l{\left(}
\def\r{\right)}

\newcommand{\be}{\begin{equation}}
\newcommand{\ee}{\end{equation}}
\newcommand{\bea}{\begin{eqnarray}}
\newcommand{\eea}{\end{eqnarray}}
\newcommand{\bg}{\begin{gather}}
\newcommand{\eg}{\end{gather}}
\newcommand{\bseq}{\begin{subequations}}
\newcommand{\eseq}{\end{subequations}}

\newcommand{\Tr}{{\rm Tr}}

\begin{document}
\baselineskip=15.5pt
\begin{titlepage}
\begin{center}
{\Large\bf   A Simple Worldsheet Black Hole }\\
\vspace{0.5cm}
{ \large
Sergei Dubovsky
}\\
\vspace{.45cm}
{\small  \textit{   Center for Cosmology and Particle Physics,\\ Department of Physics,
      New York University\\
      New York, NY, 10003, USA}}\\ 
      \vspace{.1cm}
\end{center}
\begin{center}
\begin{abstract}


  We study worldsheet theory of confining strings in  two-dimensional massive adjoint QCD. Similarly to confining strings in higher dimensions this theory exhibits 
 a non-trivial $S$-matrix surviving even in the strict planar limit.  In the process of two-particle scattering a zigzag is formed on the worldsheet. It leads to an interesting non-locality and exhibits some properties of a quantum black hole.  Ordinarily, identical quantum particles do not carry identity.
  On the worldsheet  they acquire off-shell identity due to strings attached. Identity implies complementarity.  We discuss similarities and differences of the worldsheet scattering with the $T\bar{T}$ deformation. We also propose a promising candidate for a supersymmetric model with integrable confining strings.
 
%
%
%
%
%
  
\end{abstract}
\end{center}
\end{titlepage}
\tableofcontents
\newpage
\section{Introduction}
Constructng a theory of confining strings remains an interesting challenge. Large $N$ gauge theories are closely related to gravity, with AdS/CFT  \cite{Maldacena:1997re,Gubser:1998bc,Witten:1998qj} 
 providing a concrete and spectacular example of this connection for conformal theories. The BFSS matrix model \cite{Banks:1996vh} and the matrix model for $c=1$ strings \cite{Kazakov:2000pm}
 are other notable examples.
Understanding the confining case is likely to give a further insight into gravitational dynamics.

Recent progress in understanding  of confining strings \cite{Dubovsky:2013gi,Dubovsky:2014fma,Dubovsky:2015zey,Dubovsky:2016cog}
 is to large extent related to the identification of a new nice observable to focus on---the worldsheet $S$-matrix \cite{Dubovsky:2012sh}
 (for other recent developments see, e.g., \cite{Hellerman:2013kba,Hellerman:2014cba,Caron-Huot:2016icg,Sever:2017ylk,Sonnenschein:2017ylo,Sonnenschein:2018aqf}).
 In the planar limit the worldsheet dynamics decouples from the bulk, and the worldsheet scattering is  described by a UV complete unitary two-dimensional theory. There is a number of indications that this theory exhibits certain gravitational features rather than being just a conventional local quantum field theory. If this expectation is confirmed, it provides another concrete link between large $N$ gauge theories and gravity.
 
 A somewhat surprising property of the worldsheet $S$-matrix is that it stays non-trivial even in the strict planar limit, $N=\infty$, when the bulk $S$-matrix turns free. This is an exact analogue of what happens in the critical string theory, where in the free limit, $g_s=0$, the worldsheet scattering remains non-trivial and describes an integrable model of two-dimensional gravity \cite{Dubovsky:2012wk}.

In three and four dimensions, $D=3,4$, a wealth of information about the worldsheet scattering can be extracted from  lattice studies of confining flux tubes (\cite{Meyer:2003wx,Lucini:2010nv,Athenodorou:2010cs,Athenodorou:2011rx,Athenodorou:2016kpd,Athenodorou:2016ebg,Athenodorou:2017cmw}, see \cite{Teper:2009uf,Lucini:2012gg} for reviews).
 In the present paper we turn to the two-dimensional case, $D=2$, where the analytical access 
to the worldsheet dynamics should be easier to achieve.

The study of two-dimensional QCD has a long and fascinating history. In the pure Yang--Mills case neither the ``bulk"\footnote{Note that in $D=2$ the ``bulk" theory lives in the same number of dimensions as the worlsheet one.} nor the worldsheet theory carry local degrees of freedom. The partition function and the Wilson loop corelators of this topological theory can be calculated exactly even at finite $N$  \cite{Migdal:1975zg,Kazakov:1980zi,Kazakov:1980zj,Rusakov:1990rs,Witten:1991we}. The large $N$ expansion of the exact answer can be explicitly recast in the form of perturbative string series \cite{Gross:1992tu,Gross:1993hu,Cordes:1994fc}.

Introducing quarks in the fundamental representation brings in local dynamics and results in the 't Hooft model, where one can solve for the spectrum of mesons in the planar limit \cite{tHooft:1974pnl}.
However, fundamental quarks do not respect the center symmetry, allowing for the confining strings to be broken. Hence, to study the  dynamics of confining strings it is natural to introduce adjoint quarks instead \cite{Dalley:1992yy,Bhanot:1993xp,Kutasov:1993gq}. Most likely, the theory is no longer solvable, but many details about its spectrum can still be extracted semianalytically using the light cone quantization (for a recent update see \cite{Katz:2013qua}). 

Unfortunately, so far not much wisdom has been gained from these results as far as the dynamics of confining strings in higher dimensions is concerned. In the present paper we propose that a promising remedy for this
 is  to focus again on the worldsheet scattering. Unlike a bulk theory, the string worldsheet is always two-dimensional, so it is plausible that many properties of the worldsheet $S$-matrix at $D=2$ can be directly lifted to $D=3$ and $D=4$. 

We take quarks to be Majorana fermions, which proved to be the simplest case in the studies of the hadron spectrum \cite{Dalley:1992yy}.
A peculiarity of the $D=2$ case is that for massless quarks the color charge is shielded rather than confined
\cite{Kutasov:1994xq,Gross:1995bp}. Hence we are led to consider the setup where fermions are massive, so that the action is 
\be
\label{action}
S= \int d^2\sigma\Tr\left\{-{1\over 2g^2} F^{\mu\nu}F_{\mu\nu}+\bar\psi(i\gamma^\mu\nabla^{(ad)}_\mu-m)\psi\right\}\;.
\ee 
In the present paper we restrict to the heavy mass case,
\be
\label{heavy}
m^2\gg g^2N\;,
\ee
when 
 the theory is  amenable to a straightforward perturbative expansion. 
 It will be interesting to understand the light mass regime as well. Conventionally it is studied using the bosonization technique \cite{Coleman:1975pw,Coleman:1976uz,Armoni:1997ki}.
Unfortunately, it does not appear to be useful for our purposes. The bosonic description is weakly coupled at $m^2\ll g^2$, so that the quark mass vanishes in the planar limit, and one does not get a description of the confining phase at $N=\infty$. For instance, the string tension calculated in \cite{Armoni:1997ki} vanishes in the 'tHooft limit.

The rest of the paper is organized as follows. In section~\ref{sec:theta} we describe the setup in detail.
We explain that  the string worldsheet arises as a subsector in the discrete $\theta$-vacuum present in QCD$_2$ \cite{Witten:1978ka}.  Similarly to higher dimensions this subsector decouples from the rest of the theory in the planar limit but retains a non-trivial scattering. 
$\theta$-vacua in QCD$_2$ are discrete analogs of constant electric field backgrounds studied in the Abelian massive Schwinger model by Coleman \cite{Coleman:1976uz}. We review the Hamiltonian description of these vacua and the corresponding one-particle excitations---``the free quarks".

In section~\ref{sec:2p} we study two-particle worldsheet excitations. We encounter an interesting puzzle here. Namely, naively the number of two-particle excitations on the worldsheet does not match the expectation based on the one-particle spectrum---apparently there are ``extra" two-particle excitations. These excitations have a simple string interpretation  and somewhat related to  ``half-asymptotic states" described by Coleman in the  massive Schwinger model at $\theta=\pi$ (see also \cite{Shankar:2005du} for a  detailed discussion). However, a careful examination of the two-particle Hamiltonian and of the resulting scattering reveals that the extra states are an illusion as far as the on-shell scattering data is concerned. Nevertheless, as a price for resolving this apparent contradiction the worldsheet theory exhibits an interesting non-locality.
In section~\ref{sec:BH} we discuss some of the implications of this non-locality. We argue that two-particle excitations on the worldsheet exhibit some features expected from a quantum black hole. In particular this system provides a specific realization of  complementarity.

In  section~\ref{sec:TT} we compare worldsheet scattering to the gravitational dressing introduced in \cite{Dubovsky:2013ira} and recasted recently  in the operator formalism as a $T\bar{T}$ deformation  \cite{Caselle:2013dra,Smirnov:2016lqw,Cavaglia:2016oda}, building up on \cite{Zamolodchikov:2004ce}.
We describe some tantalizing similarities but also some differences.
 We also point out that in the light mass regime a particularly interesting point is 
\[
m^2={g^2N\over \pi}\;.
\]
The theory (\ref{action}) is known to be supersymmetric at this point \cite{Kutasov:1993gq}. The supersymmetry is spontaneously broken on the worldsheet with the free quark turning into a massless 
goldstino. This theory provides an interesting candidate for an integrable confining string model.

We conclude in section~\ref{sec:last}.
Appendices \ref{app:conv} and \ref{app:dull} contain various conventions and technical details of our  calculations.
\section{String Worldsheet  as a $\theta$-vacuum in QCD$_2$}
\label{sec:theta}
Confining flux tubes in dimensions $D=3,4$ are easily visualized as  lower dimensional dynamical defects embedded in a higher dimensional bulk. The notion of a worldsheet theory at $D=2$ may appear somewhat confusing, given that it has the same dimensionality as  the full ``bulk" theory.
 So let us start by explaining the precise meaning of the worldsheet theory at $D=2$.

\begin{figure}[t!]
  \begin{center}
        \includegraphics[height=6cm]{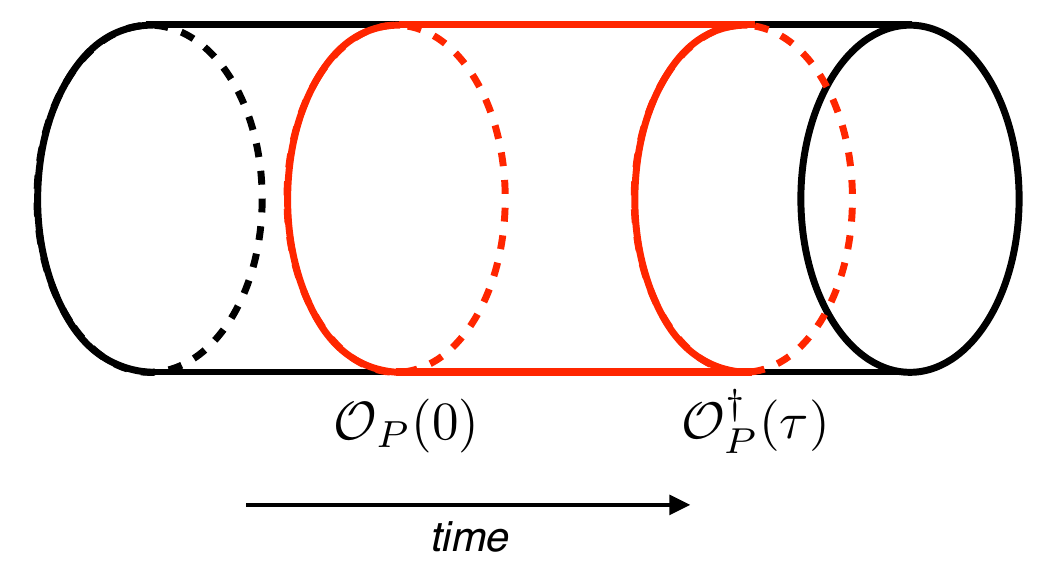} 
           \caption{ A long string wound around the cylinder is created by a Polyakov loop ${\cal O}_P$. In lattice simulations a finite volume spectrum of the worldsheet theory is determined from the exponential falloff of two-point correlators, $\langle {\cal O}^\dagger_P(\tau) {\cal O}_P(0) =\sum_n e^{-E_n\tau}$.  }
        \label{fig:loops}
    \end{center}
\end{figure}

 Note first that at $D=2$ the confinement is present even in a classical Abelian theory. Hence, one may expect that the presence of a confining string should be equivalent to a background of a constant (chromo)electric field generated between confined (chromo)electric charges. To make this precise it is instructive to recall how  flux tubes are observed in the lattice simulations in $D=3,4$ gluodynamics\cite{Teper:2009uf}. There one considers a compactification on a cylinder and measures a two-point correlator of  Polyakov loops,
 \[
 {\cal O}_{\cal P}=\Tr Pe^{ i\int_{\cal P}d\sigma A}\;,
 \]
 where the path ${\cal P}$ winds once around the spatial circle, and the trace is taken in the fundamental representation. This operator creates a wound string state and by measuring the exponential falloff of the two-point function $ \langle {\cal O}_{\cal P}^\dagger(\tau){\cal O}_{\cal P}(0)\rangle$ one determines its energy, see Fig.~\ref{fig:loops}. 
 Different worldsheet states may be probed by deforming the integration contour ${\cal P}$ in the transverse directions.
  As a consequence of the center symmetry only states with a wrapped string present contribute into this correlator. At finite $N$ the operator $ {\cal O}_{\cal P}$ may produce also a bunch of glueballs in addition to a wound string. However, this does not happen in the planar limit, so that at large $N$ this measurement allows to determine the finite volume spectrum of the worldsheet theory.
  
 Exactly the same strategy can be applied at $D=2$. Of course in this case the fixed time slices are one-dimensional, so all paths ${\cal P}$ give the same result. This agrees with the absence of local excitations on the string worldsheet in the two-dimensional pure glue theory. However, in the presence of adjoint fermions one may consider generalized Polyakov loops with additional fermion insertions,
  \be
  \label{Ofs}
 {\cal O}_{{\cal P},n}=\Tr Pe^{ i\int_{\cal P}d\sigma A}\psi(\sigma_1)\dots\psi(\sigma_n)\;,
 \ee
which create  non-trivial local excitations on the worldsheet. Upon taking the circle size  to infinity this construction matches the description of a discrete $\theta$-vacuum introduced in \cite{Witten:1978ka}.
 
The only distinction between the $\theta$-vacuum and what we call the string worldsheet is that the latter describes a subsector of the former spanned by  states created by single trace operators of the form (\ref{Ofs}). A generic excitation of the $\theta$-vacuum is created by a multitrace operator constructed as a product of (\ref{Ofs}) and a set of topologically trivial Wilson loops with fermion insertions producing ``glueball" states in addition to a long string. In the planar limit glueball states decouple.

Dynamics in the $\theta$-vacuum is conveniently described by the Hamiltonian formalism presented in
 \cite{Witten:1978ka}, which generalizes Coleman's treatment of the massive Schwinger model \cite{Coleman:1976uz}. Instead of a circle compactification it is more convenient to work on an interval ${\cal I}=[-L,L]$ with infinitely heavy fundamental and antifundamental charges at the end-points. In the limit $L\to \infty$ there is no distinction between a circle and an interval anyway.
  
To construct the canonical formalism one fixes the axial gauge 
\[
A^a_1=0\;.
\]
 The remaining components of the gauge field $A^a_0$ are non-dynamical, so one can solve for them from the Gauss' law and plug the result back in the action. The solution for the chromoelectric field takes the form
 \be
\label{Ea}
E^a\equiv \d_\sigma A_0^a={g^2\over 2}\int d\sigma' \epsilon(\sigma-\sigma')\l\rho^a(\sigma')+\rho_{bd}^a(\sigma')\r\;,
\ee
 where
$\epsilon(\sigma)$ is the sign function. Here 
\be
\label{rho}
\rho^a=-{i\over 2}\psi^{Tb}\psi^cf^{abc}\;.
\ee
 is the charge density of dynamical fermions and  $\rho_{bd}^a$ accounts for the presence of heavy charges at the boundaries,
 \[
\rho_{bd}^a=T^a\delta(\sigma-L)+\bar{T}^a\delta(\sigma+L)\;.
\]
 $T^a$ and $\bar{T}^a$ are  generators of the fundamental and antifundamental representations acting on the boundary charges. Importantly, these are quantum mechanical operators rather than $c$-numbers. They may be thought to describe dynamics of internal color degrees of freedom which survive even in the limit when the mass of the boundary charges is taken to infinity. 
 
 To describe physical states one  starts with an enlarged Hilbert space, 
 \[
 {\cal H}_0=V\otimes{\cal H}_f\otimes \bar{ V}\;,
 \] 
 where ${\cal H}_f$ is the Hilbert space of free dynamical quarks $\psi^a$, and $V$, $\bar{V}$ are vector spaces describing the states of boundary fundamental and antifundamental charges. Physical states 
 $|ph\rangle$ 
 are the states of ${\cal H}_0$ which do not carry color, {\it i.e.}, they are annihilated by  total color charges
\be
\label{Qa}
Q^a|ph\rangle=\l T^a+\bar{T}^a+\int \rho^a\r|ph\rangle=0\;.
\ee
 The total Hamiltonian is a sum of a free fermion Hamiltonian and of the energy stored in the chromoelectric field,
 \be
 \label{Ham}
 H=\int d\sigma \l{1\over 2g^2}(E^a)^2+{1\over 2}\bar\psi^a(-i\gamma^1\d_\sigma+m)\psi^a\r\;,
 \ee
 where the latter may be presented as the sum of three terms,
 \[
\int d\sigma {1\over 2g^2}(E^a)^2=H_0+H_2+H_4\;.
 \]
 The first term
 \be
 \label{H0}
 H_0=2L {g^2}{{\cal C}_2\over 2}
 \ee
  is the vacuum energy responsible for the classical string tension with ${\cal C}_2$ standing for the quadratic Casimir in the fundamental representation,
  \[
  {\cal C}_2=(T^a)^2={N^2-1\over2 N}\approx {N\over 2}\;.
  \]
 The second term describes  interaction of dynamical quarks with  the chromoelectric field 
 produced by the boundary charges,
  \be
  \label{H2}
  H_2={g^2\over 2}(\bar{T}^a-{T}^a)\int d\sigma \sigma\rho^a(\sigma)\;,
  \ee
  Naively this term is not translationally invariant. However, after multiplying the charge neutrality condition (\ref{Qa}) by $(\bar{T}^a-{T}^a)$ one finds that the $H_2$ variation under translations vanishes in the physical subspace of ${\cal H}_0$.
  
  The last term
  \be
  \label{H4}
  H_4=-{g^2\over 4}\int d\sigma d\sigma'|\sigma-\sigma'|
\rho^a(\sigma)\rho^a(\sigma')
  \ee
  describes non-Abelian Coulomb interactions between quarks due to  gluon exchange.
  
  To conclude the review of the $\theta$-vacuum let us describe the ground state of a long string and one-particle excitations on its worldsheet. In the heavy mass limit the theory is weakly coupled so the mixing between states with different number of quarks (``partons") is suppressed.
  To avoid an insane proliferation of indices it is convenient to think of states in ${\cal H}_0$ as matrices acting in the fundamental representation with entries taking values in the quark Fock space ${\cal H}_{free}$.
  
  Then, for a large quark mass, the normalized physical ground state $|0\rangle$ is well approximated by a single color singlet state without parton excitations,
  \[
  |0\rangle ={1\over\sqrt{N}}|0\rangle_F\otimes{\mathbf 1}\;,
  \]
  where $|0\rangle_F$ is the Fock vacuum and ${\mathbf 1}$ is the unit operator acting in the fundamental representation. The string tension $\ell_s^{-2}$ is determined mostly by the classical vacuum energy  (\ref{H0}),
  \be
  \label{tension}
  \ell_s^{-2}={g^2N\over 4}\;.
  \ee
  
\begin{figure}[t!]
  \begin{center}
        \includegraphics[width=10cm]{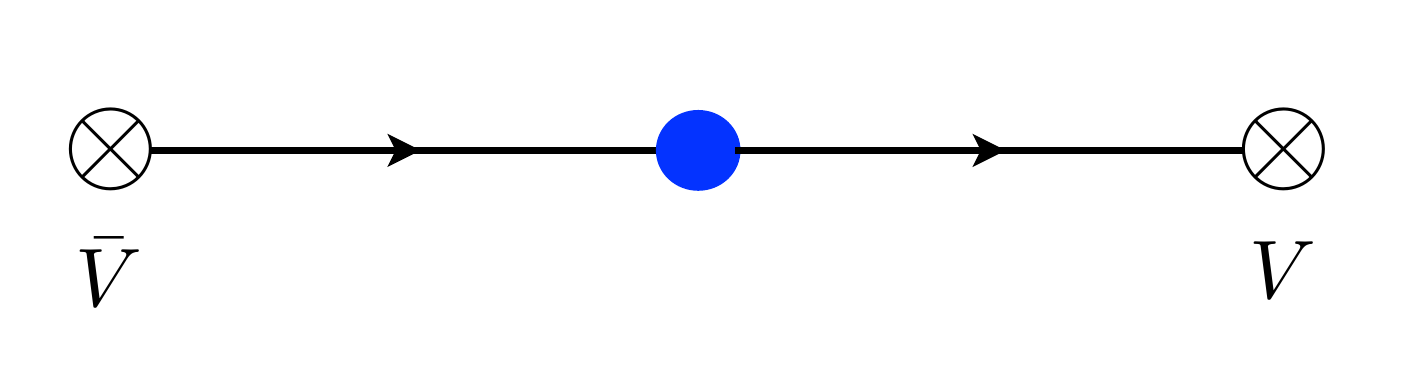} 
           \caption{ A one-particle state on the worldsheet may be represented  as a free quark with two conjugate flux lines emanating in opposite directions and terminating at the boundary charges at infinity.  }
        \label{fig:1p}
    \end{center}
\end{figure}

 Similarly, one-particle states are well approximated by color singlet  one-parton states in ${\cal H}_0$.
 For any value of the spatial momentum $k$ there is a single such state,
\be
 |k\rangle=N_1|k,a\rangle \otimes T^a\;,
 \ee   
  where $|k,a\rangle$ is a one-particle state in ${\cal H}_{free}$,
  \[
  |k,a\rangle=\beta_k^{a\dagger}|0\rangle_F
  \]
  and the normalization factor 
  \[
 N_1=(N{\cal C}_2)^{-1/2}\approx {\sqrt{2}\over N}\;,
  \]
  is fixed from 
 \[
\langle q |k\rangle=|N_1|^2\delta(k-q)\Tr T^{\dagger a}T^a=\delta(k-q)
\]
These states describe a single massive particle on the worldsheet---the ``free quark". Pictorially it can be represented as an adjoint charge with two conjugate flux lines attached. The flux lines
stretch into opposite directions and terminate at the boundary charges,
Fig.~\ref{fig:1p}. Clearly, these states correspond to operators (\ref{Ofs}) with a single fermion insertion.

\section{Two-Particle States on the Worldsheet}
\label{sec:2p}
\subsection{Distinguishable Particles}
To confirm that the worldsheet scattering has a non-trivial planar limit we need to inspect  properties of two-particle states. In order to avoid additional complications arising in the presence of identical particles 
it is instructive to modify the model by considering its two flavor version,
\[
\psi^a\to(\psi^a_1,\psi^a_2)\;.
\]
For simplicity, we first consider only two-particle states where different parton flavors are excited. When the dust settles, it will be straightforward to extend the discussion to the states with identical particles.

By inspecting color singlet states in ${\cal H}_0$  we immediately encounter the following  puzzle.
Namely, it is easy to see that there are {\it three} families of physical two-parton states in ${\cal H}_0$,
\begin{gather}
\label{m}
|k_1,k_2\rangle_m\approx N^{-3/2}|k_1,k_2,a_1,a_2\rangle_F \otimes \delta^{a_1a_2}{\bf 1}\\
\label{L}
|k_1,k_2\rangle_L\approx2N^{-3/2}|k_1,k_2,a_1,a_2\rangle_F \otimes T^{a_1}T^{a_2}\\
\label{R}
|k_1,k_2\rangle_R\approx2N^{-3/2}|k_1,k_2,a_1,a_2\rangle_F \otimes T^{a_2}T^{a_1}\;,
\end{gather}
where 
\[
|k_1,k_2,a_1,a_2\rangle_F=\beta_{k_1}^{a_1\dagger}\gamma_{k_2}^{a_2\dagger}|0\rangle_F\;.
\]
The states (\ref{m}), (\ref{L}), (\ref{R}) are approximate in the sense that they form an orthonormal basis only at $N=\infty$.
It is straightforward to generalize these expressions into exactly orthonormal states, but we do not need them.
These three families of states were previously identified in  \cite{Paniak:1996zn}, where their presence was also deduced from the exact answer for the expectation value of the nested fundamental and adjoint Wilson loops in the pure glue theory. The puzzle is how the existence of  three families of two-particle states compatible with a single one-particle state we previously found on the worldsheet for each flavor? 

\begin{figure}[t!]
  \begin{center}
        \includegraphics[width=10cm]{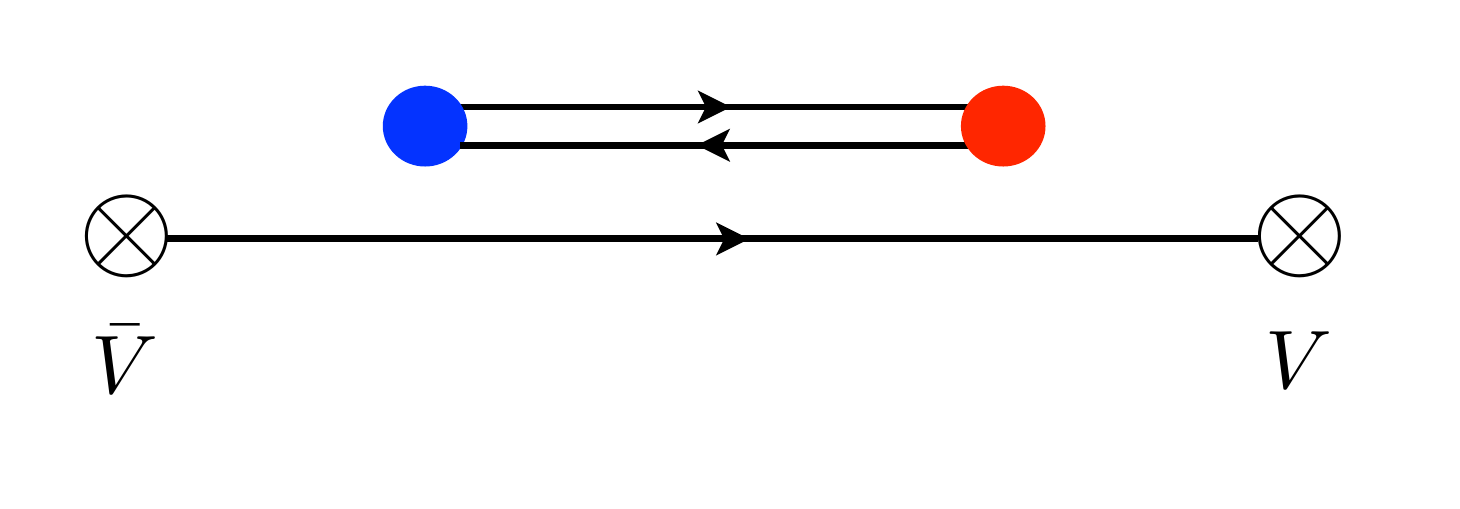} 
           \caption{ Meson states are created by  double-trace operators and decouple from worldsheet excitations in the planar limit.
             }
        \label{fig:mesons}
    \end{center}
\end{figure}

In fact, the first family of states $|k_1,k_2\rangle_m$ does not present a puzzle at all. As indicated by the contraction of color indices, these states describe bulk mesons rather than worldsheet excitations, see Fig.~\ref{fig:mesons}.
 It is straightforward to check that at large $N$ these states  decouple from the worldsheet states $|L\rangle$, $|R\rangle$. 
Indeed, note first that the gluon mediated interaction $H_4$ commutes separately with the bulk  $Q^a_0$ and the boundary $T^a+\bar{T}^a$ color charges. Meson states form a singlet  w.r.t. the boundary charge, while all worldsheet excitations are linear combinations of states in the adjoint of the boundary charge. Hence, $H_4$ alone cannot induce mixing between mesons and worldsheet excitations.

On the other hand, acting with $H_2$ on the meson state we get (restricting to a two-particle sector)
\[
H_2|m\rangle={g^2\over i N^{3/2}}\l\d \beta_{k_1}^{a_1\dagger}\gamma_{k_2}^{a_2\dagger}- \beta_{k_1}^{a_1\dagger}\d\gamma_{k_2}^{a_2\dagger}\r|0\rangle\otimes [T^{a_1},T^{a_2}]\;,
\]
which is a worldsheet state\footnote{A somewhat unconventional notation $\d \beta^{a\dagger}_k$ is explained in the Appendix~\ref{app:conv}.}. However,  in the 't Hooft limit $g^2N=const$ this result is $1/N$ suppressed compared to the unperturbed worldsheet states
$|L\rangle$, $|R\rangle$. Hence the string breaking is impossible in the planar limit, as 
it should be.

\begin{figure}[t!]
  \begin{center}
        \includegraphics[width=17cm]{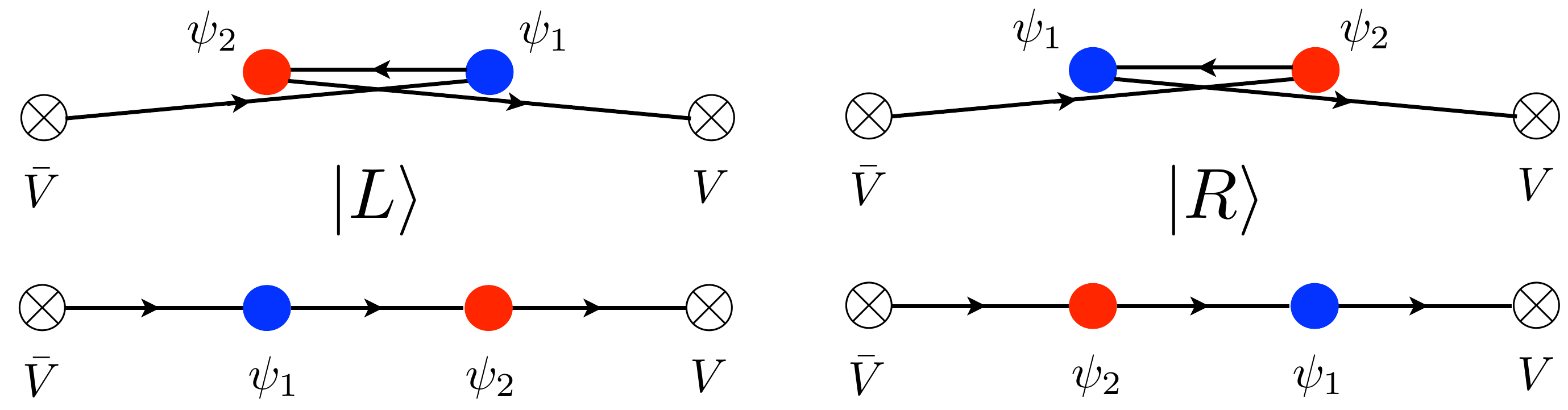} 
           \caption{  $|L\rangle$ and  $|R\rangle$ sectors arise as a consequence of the color ordering.  Note that the ordering of quarks in space does not need to agree with the color ordering. This leads to a possibility of zigzag configurations, as shown on the top.
             }
        \label{fig:2p}
    \end{center}
\end{figure}

The remaining two states do however pose a puzzle. Actually, their presence also  has a natural pictorial representation in the string language. Indeed, if one visualizes quarks as beads with confining strings (``rubber bands") stretched through them (cf. Fig.~\ref{fig:1p}) one naturally arrives at two different states corresponding to two color index contractions in (\ref{L}) and (\ref{R}), see Fig.~\ref{fig:2p}. 
Note that, as illustrated in Fig.~\ref{fig:2p}, an index contraction does not need to match the physical ordering of quarks in space---the presence of quarks allows to introduce zigzags on the Polyakov loop. However,  particles in quantum field theory typically behave differently from beads on a string. The worldsheet theory has a single one-particle state corresponding  to each of the quark flavors, so its two-particle excitations containing different quark flavors should 
also be described by a single family of scattering states. 

\begin{figure}[t!]
  \begin{center}
        \includegraphics[width=11cm]{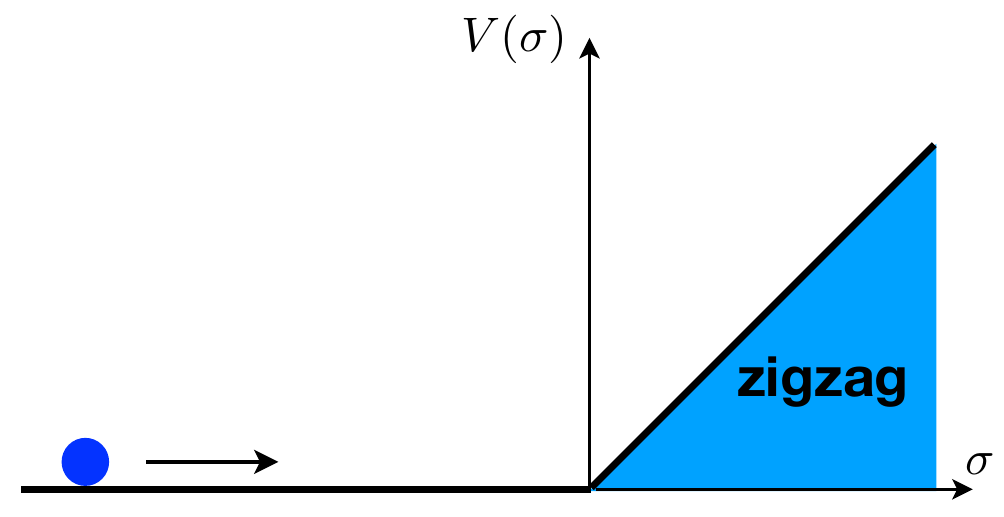} 
           \caption{ Potential describing a two-particle collision in the $|L\rangle$-sector.
             }
        \label{fig:pot}
    \end{center}
\end{figure}

To see how this apparent contradiction is resolved let us study dynamics of the $|L\rangle$, $|R\rangle$
states. Neglecting the mixing with multiparticle states this amounts to calculating matrix elements of the full Hamiltonian  (\ref{Ham}) in the  $|L\rangle$, $|R\rangle$ subspace, similarly to the Abelian case \cite{Coleman:1976uz}.
 This results in the effective center-of-mass Hamiltonian $H_{eff}$ defined by
\be
\langle k'_1,k'_2|H|k,-k\rangle=\delta(k'_1+k'_2)\langle k'|H_{eff}|k\rangle\;.
\ee
The details of the calculations are presented in the Appendix~\ref{app:dull}.
As follows from eqs. (\ref{H2fin}), (\ref{H4fin}) there, the $|L\rangle$ and  $|R\rangle$ sectors are completely decoupled, and the corresponding Hamiltonians
are
\be
\label{Heff}
\langle k'|H_{eff}^{L(R)}|k\rangle=\delta(k-k')2\omega_k-{g^2N\over 4\pi}\l{\cal U}(k,k'){{\cal P}\over (k-k')^2}\pm i\pi\delta'(k-k')\r
\ee
The function ${\cal U}(k,k')$ is given by (\ref{calU}) and it is equal to unity both in a non-relativistic regime $k,k'\ll m$ and in the ultra-relativistic one. Hence, in view of (\ref{absF}), after transitioning into the position space, each of the effective Hamiltonians (\ref{Heff}) essentially describe a free fermion moving in a potential of the form
\be
\label{potential}
V^{L(R)}(\sigma)={g^2N\over 4}(|\sigma|\pm\sigma)\;,
\ee
see Fig.~\ref{fig:pot}. This potential does give rise to a non-trivial scattering phase shift,  confirming that the world-sheet $S$-matrix is indeed non-trivial in the strict planar limit.

Furthermore, we see that the ``beads on a rubber band" picture turns out correct at the end of the day. In the two-to-two scattering of two ``free" quarks of different flavors they always end up reflecting from each other, as a consequence of the semiconfining  potential (\ref{potential}). The linearly growing segment of this potential is due to a confining string formed during the zigzag period of the collision shown on the top panel of Fig.~\ref{fig:2p}. As a consistency check, note that the slope of the potential (\ref{potential}) is twice larger than the tension of the fundamental string  (\ref{tension}) in agreement with the zigzag interpretation.

On the other hand, thanks to the  semiconfining nature of the potential (\ref{potential}),  the standard  counting of asymptotic states holds as well. Namely, the $|L\rangle$-sector describes in-states where the $\psi_1$ quarks enters the collision from  the left, and $\psi_2$ enters from the right. Similarly, for out-states in the $|L\rangle$-sector  $\psi_1$ exits the collision on the left, while $\psi_2$ exits on the right. The $|R\rangle$-sector
describes the complementary kinematics with left and right interchanged.

\begin{figure}[t!]
  \begin{center}
        \includegraphics[height=6cm]{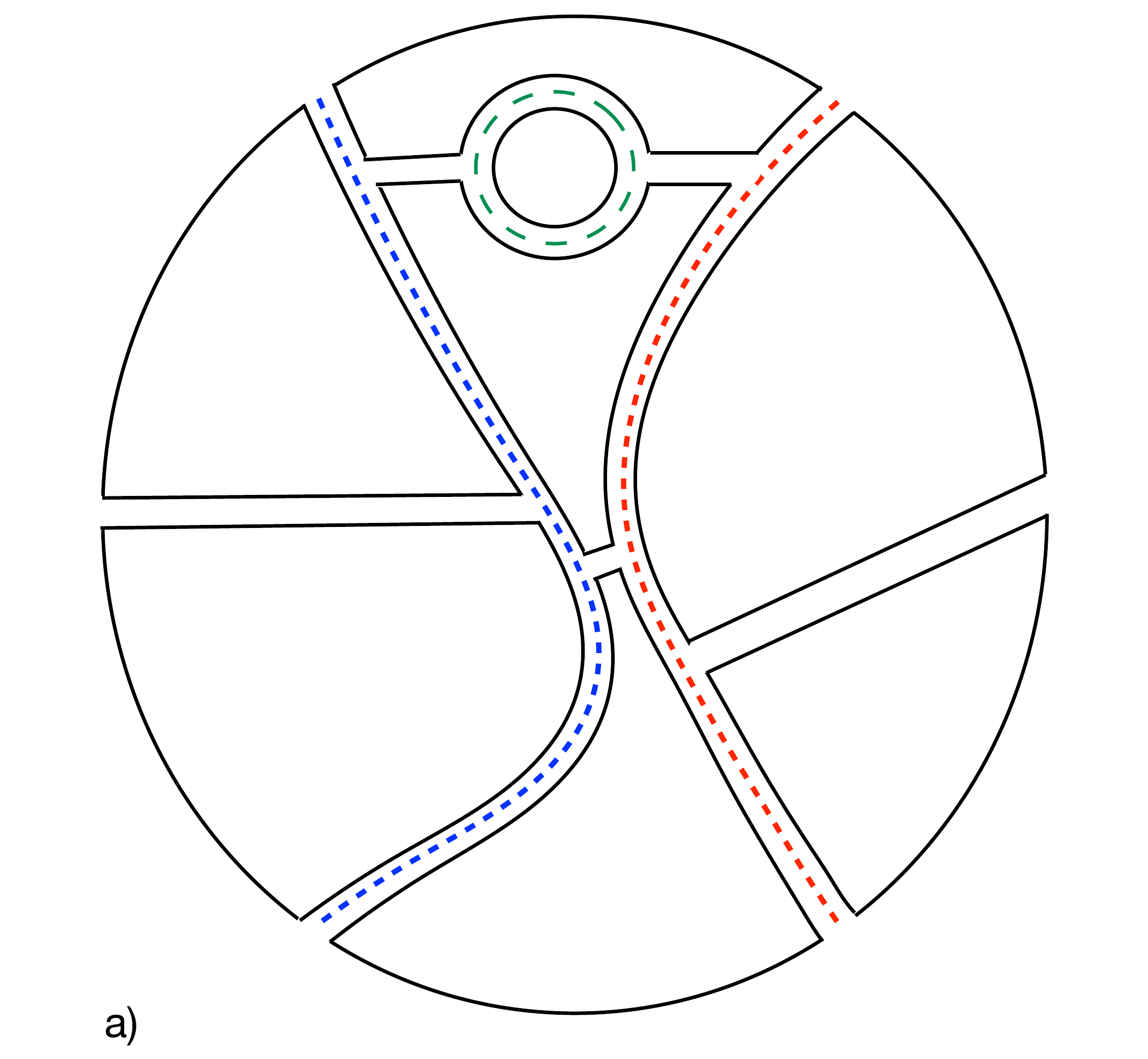} \;\;\;\;\;
          \includegraphics[height=6cm]{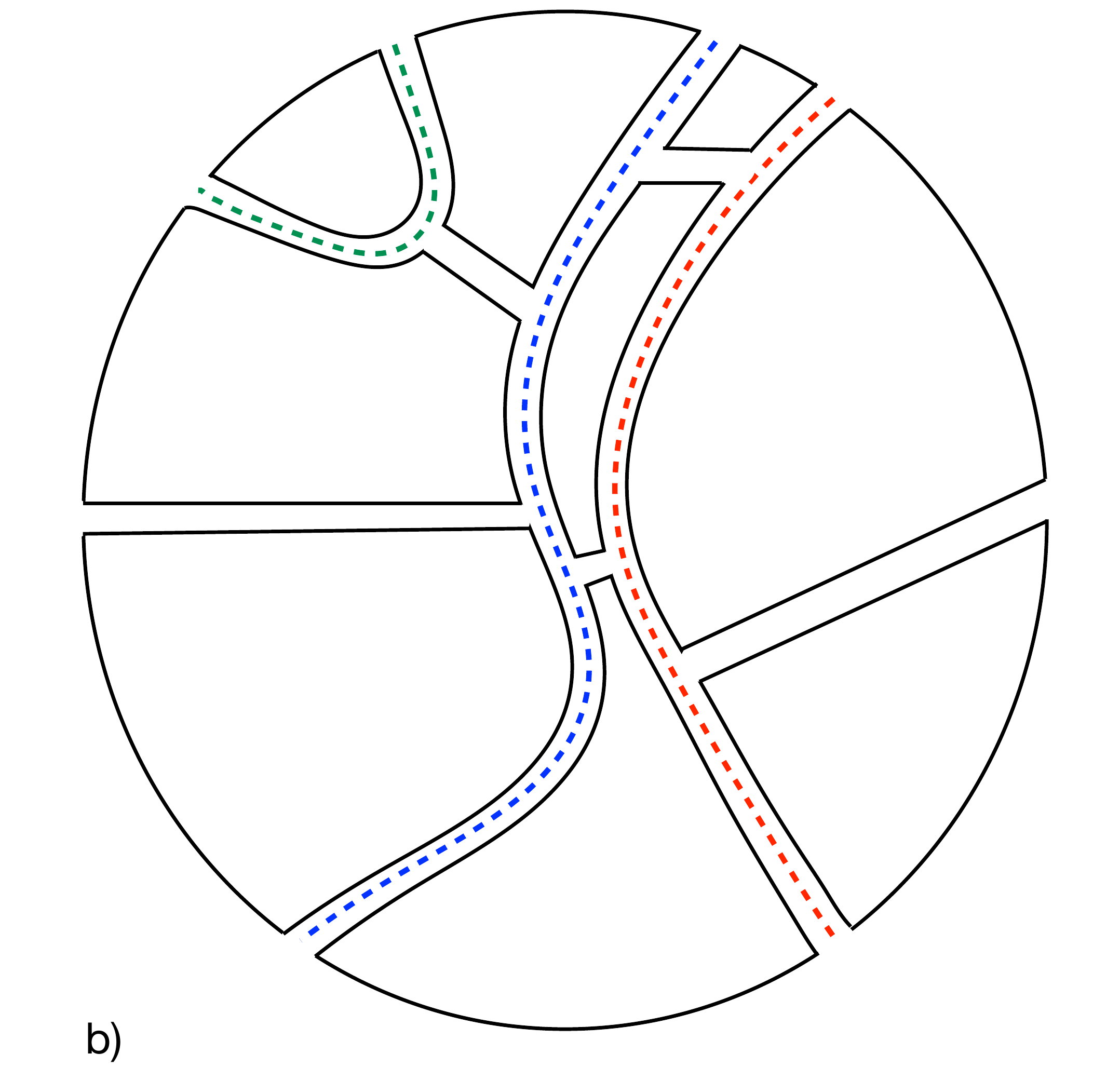} 
           \caption{a) As a consequence of planarity two-particle collisions on the worldsheet reduce to pure reflection. b) This is no longer the case for multiparticle processes.  However, planarity still highly restricts possible flavor orderings at infinity. On both panels dashed colored lines show the flow of quark flavors. Empty double lines show gluons exchange.
             }
        \label{fig:plan}
    \end{center}
\end{figure}

The absence of transitions in  the two-to-two scattering remains true also at higher orders in the perturbation theory, with multiparton intermediate states taken into account. The fastest way to see this is to use a covariant description of the $\theta$-vacuum \cite{Witten:1978ka}. Namely, adding auxiliary charges at infinity is equivalent to calculating 
a path integral on a plane in the presence of a large circular Wilson line,
\be
\label{covariant}
Z_\theta[\eta]=\int {\cal D}\psi{\cal D}\bar{\psi}{\cal D} A_\mu e^{iS_{QCD}}\Tr P e^{i\oint d\sigma^\mu A_\mu+i\oint ds \eta\psi}\;,
\ee
where we introduced a fermionic source $\eta(\sigma)$ along the Wilson line, so that $ Z_\theta[\eta]$ becomes a generating functional for worldsheet scattering amplitudes.
As usual, in the 't Hooft limit only the planar diagrams survive, so that the two-to-two scattering of different flavors reduces to the total reflection, see Fig.~\ref{fig:plan}a). For multiparticle processes of the type $\psi_1\psi_2\to many$ the $\psi_1$-flavor does not need to reflect back any more, see e.g. Fig.~\ref{fig:plan}b). However, planarity still imposes strong restrictions on possible orderings of the incoming and outgoing flavors along the circle at infinity.

\subsection{Identical Particles}
To conclude the discussion of the two-to-two scattering on the worldsheet let us discuss what happens when identical particles are present in the initial/final states. We turn back to the single flavor case and consider the $\psi\psi\to\psi\psi$  scattering.
In this case we have a single family of the worldsheet excitations (in addition to mesons, which again decouple in the planar limit), which can be chosen in the form
\be
\label{ident}
|k_1,k_2\rangle\approx2N^{-3/2}|k_1,k_2,a_1,a_2\rangle_F \otimes T^{a_1}T^{a_2}\;.
\ee
However, the apparent extra state is still present. Indeed, ordinarily
two-particle states for identical particles can be uniquely mapped into a half-plane, parametrized 
 by a total momentum $k_t\in(-\infty,\infty)$,
and a positive relative momentum $k_r\geq0$. The states (\ref{ident}) cover the whole plane instead, 
\[
k_1,k_2\in(-\infty,\infty)\;.
\]
  Another way to see that particles apparently acquired  identity is to calculate the scalar product of the states (\ref{ident}). Ignoring $1/N$ corrections one finds\footnote{Note that in our notations $\langle k'_1,k'_2|= 2N^{-3/2}\langle 0|\beta^{a_2}_{k'_2}\beta^{a_1}_{k'_1}\otimes T^{a_2}T^{a_1}$.},
 \[
 \langle k'_1,k'_2|k_1,k_2\rangle=\delta(k_1-k'_1)\delta(k_2-k'_2)\;,
 \]
so that the exchange term is missing. So the states (\ref{ident}) seem to describe a system of two distinguishable particles---a string attached to each of the particles allows to tell them apart.

This is a problem, because quantum particles are not entitled for an identity.
To see that the very same string saves the day as before, let us calculate the effective Hamiltonian in this case.
The relevant matrix elements are again presented in the Appendix~\ref{app:dull}.  They are very similar to what we previously found for two flavors. The only novelty is that in addition to an interaction with the background field
(\ref{H2id}) and a transition due to gluon exchange (\ref{H4tr}),  an  annihilation amplitude (\ref{H4an}) is now present. The resulting effective Hamiltonian is
\be
\label{Heffid}
\langle k'|H_{eff}|k\rangle=\delta(k-k')2\omega_k-{g^2N\over 4\pi}\l{{\cal U}(k,k')}{{\cal P}\over (k-k')^2}+ i\pi\delta'(k-k')-{m^2\over 4\omega_{k_1}^2\omega_{k'_1}^2}\r\;.
\ee
The corresponding potential in the position space is 
\be
\label{Vid}
V(\sigma)={g^2N\over 4}(|\sigma|+\sigma)+V_{ann}(\sigma)\;,
\ee
where the annihilation contribution is suppressed at high energies and reduces to the contact interaction
\be
\label{Vann}
V_{ann}(\sigma)\simeq {g^2 N\over 8 m^2}\delta(\sigma)
\ee
in the non-relativistic regime.

We see that the story is essentially the same as in the two-flavor case.
The effective Hamiltonian $H_{eff}$ describes incoming waves with $k>0$ and scattered waves with $k<0$.
 The confining string, which caused the problem to start with, also provides the cure. Confining dynamics eliminates extra 
scattering states, so that counting of the on-shell scattering modes works exactly as expected for  identical particles.
\section{Complementarity from Identity}
\label{sec:BH}
We see that the apparent contradiction with the counting of states gets happily resolved as far as the on-shell data is concerned. However, as we will discuss now, the resolution comes at a price. Local
off-shell observables need to be sacrificed in the worldsheet theory. This is consistent with the expected emergent gravitational dynamics on the worldsheet. In fact, as we argue now, a two-particle sector we just described exhibits certain features expected from a quantum back hole. In particular, non-locality exhibited by this system provides a specific scenario for the black hole complementarity. This mechanism for complementarity can be summarized in a nutshell by the slogan which went into the title of the present section. 

\begin{figure}[t!]
  \begin{center}
        \includegraphics[height=6cm]{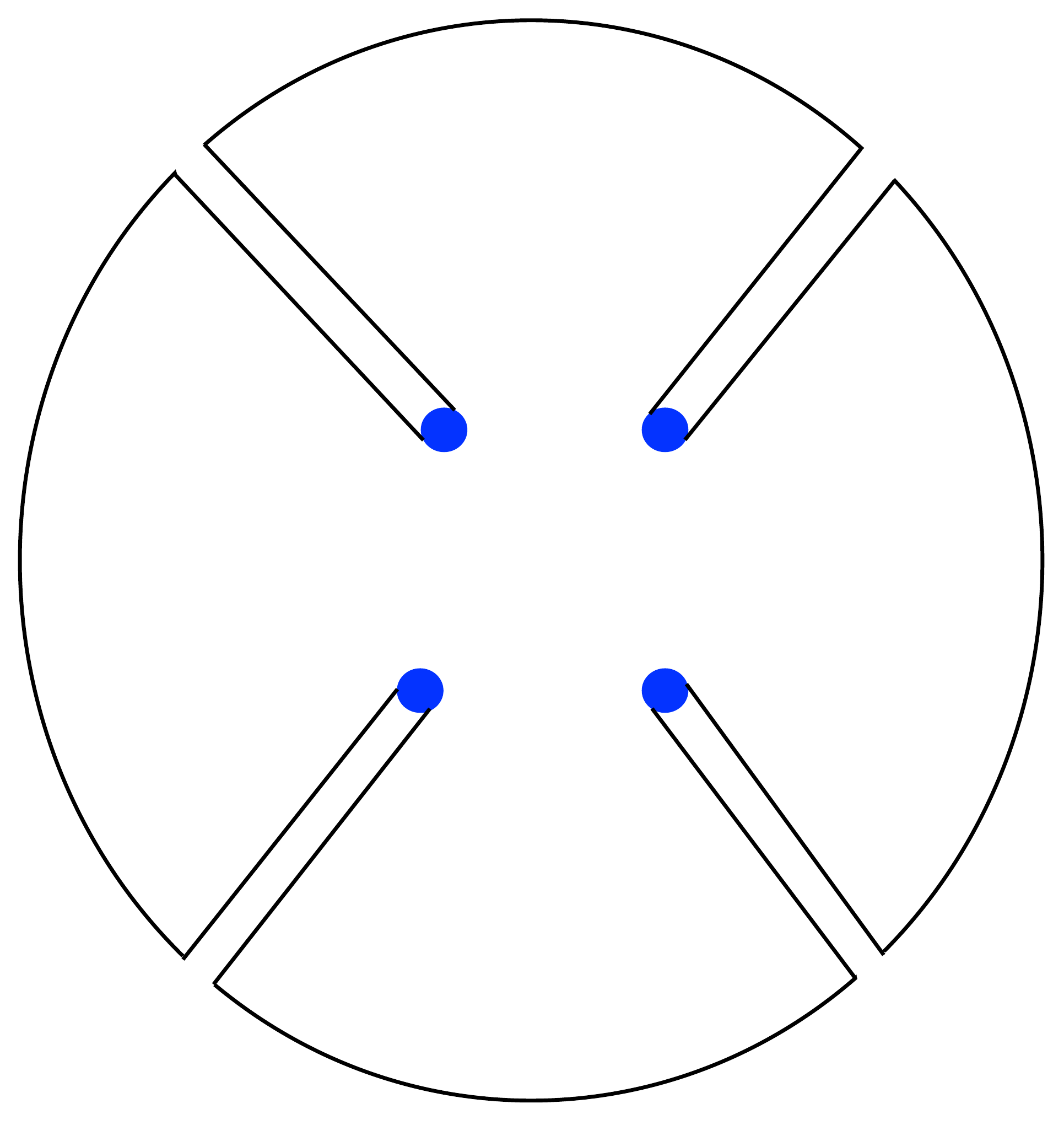}        
           \caption{Deformed Wilson lines with quark insertions immersed into the bulk provide semilocal observables for the worldsheet theory.}
        \label{fig:deep}
    \end{center}
\end{figure}

One may be surprised to encounter a non-local sector in what initially looked as a pretty conventional and a very simple local quantum field theory. Clearly, non-locality came across as a result of the $N=\infty$ limit. 
Its origin is  clearly seen in the covariant definition of the worldsheet amplitudes given by the path integral (\ref{covariant}). Lacking multitrace operators,
 the closest one may come to probe an interior of the worldsheet is to deform the Wilson line at infinity so that fermion insertions immerse inside the bulk, see Fig.~\ref{fig:deep}. This construction is reminiscent of the 
 gravitational Wilson line proposal to define semilocal observables in gravity (see, e.g., \cite{Anand:2017dav} for a recent discussion). However, due to a string connecting the bulk probe to the boundary Wilson loop at infinity it clearly fails to provide truly local observables.

 Also note that in the conventional AdS/CFT context $1/N$ corrections lead to  a {\it breakdown} of the bulk locality, so the statement above that  non-locality of the worldsheet theory is associated 
 to taking the planar limit may sound surprising. There is no contradiction here. The planar limit is required  to define the worldsheet theory on the first hand. Finite $N$ corrections do make it more non-local (and non-unitary).
 For instance, they introduce mixing with meson states, which violate worldsheet unitarity, and may be considered as the creation of baby universes (c.f. \cite{Rubakov:1995zb}). Of course, these very states also allow to define
 local observables in the full bulk theory.
 
 Coming back to specific manifestations of non-locality in the two-particle sector, these are quite explicit.  Even though on-shell scattering data matches the conventional quantum field theory expectations, particles do remain distinguishable  in the bulk. There is a string attached to each particle, and even if the flavor may be the same, the string provides an off-shell  label for individual particles.
 
 To avoid a contradiction, there is no single wave function which covers the whole space. For instance, in the two flavor case we may either provide a wave function description for the situation
 when $\psi_1$ comes from the left, covers some (energy-dependent) region in the bulk and exits to the left (the $|L\rangle$ state) or its image under reflections ($|R\rangle$), but not for the both. It is important to stress that this phenomenon is different from the conventional inability to describe relativistic quantum field theory with a non-relativistic wave function. Here the issue is present  arbitrarily deep in the non-relativistic regime, and the effect is not suppressed by a particle velocity. 
 
 As an example, one may consider a linear superposition 
 \[
| \psi\rangle=|L\rangle+|R\rangle\;.
\]
In the conventional quantum mechanics there is no problem to describe a state like this by a  wave function in the position space. The wave function  describes two wave packets, which are well separated initially and interfere at intermediate times. No local description like this is available for superpositions  of two-particle states on the worldsheet, no matter how small the velocities are.

\begin{figure}[t!]
  \begin{center}
        \includegraphics[height=6cm]{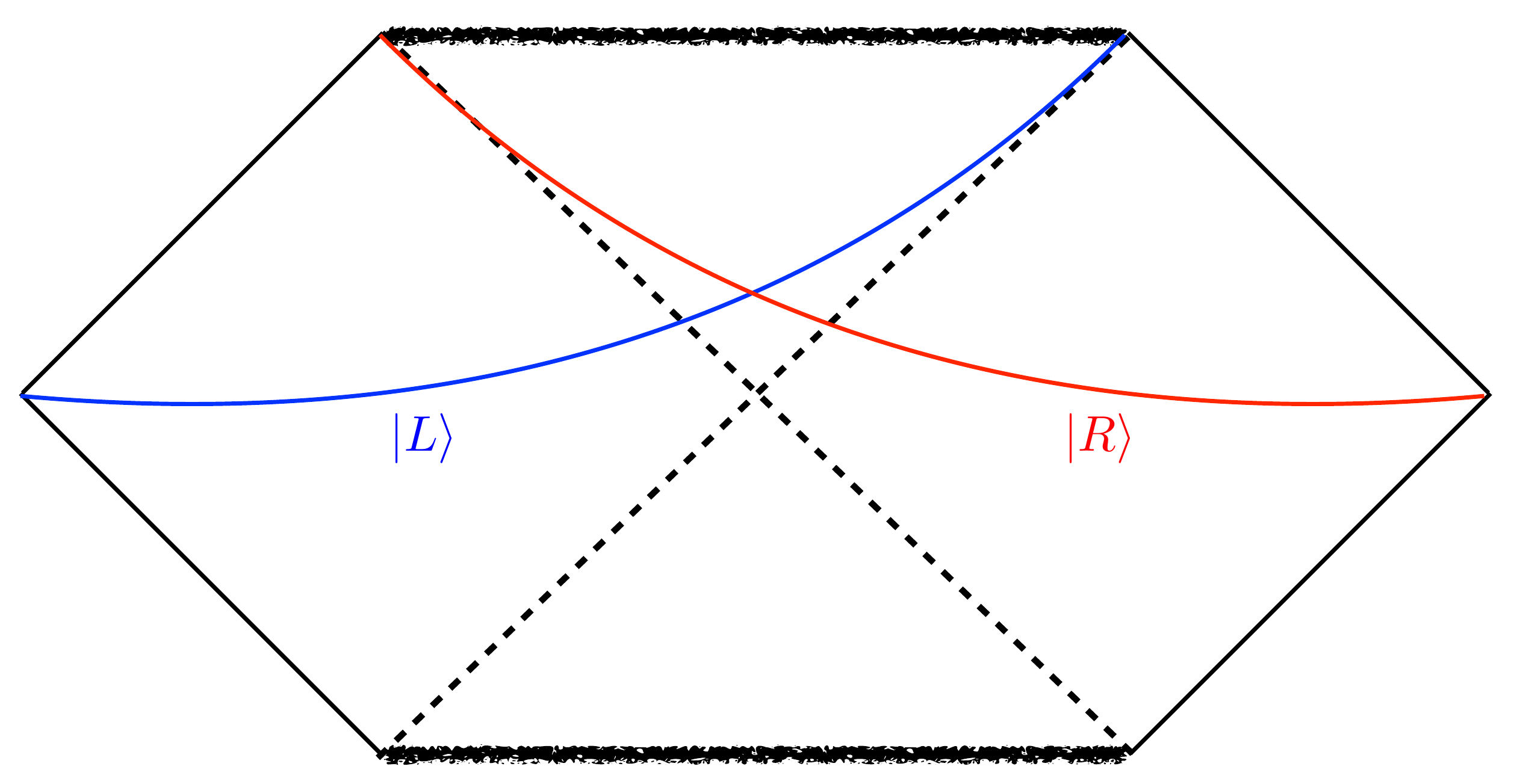}        
           \caption{$|L\rangle$ and $|R\rangle$ sectors correspond to wave functions on the two complementary spatial slices in the two-sided black hole interpretation of the zigzag collision.}
        \label{fig:Penrose}
    \end{center}
\end{figure}

At this point it is apropriate to draw an analogy between the present system and a two-sided black hole in the two-dimensional Minkowski space. There is an obvious parallel between states on the two space-like slices shown in Fig.~\ref{fig:Penrose} and  $|L\rangle$, $|R\rangle$ states. Zigzag region of the collision provides a counterpart of the black hole interior in this analogy.  

Given that the scattering is completely reflectional,
one may expect a $\psi_2$ particle to serve as an impenetrable barrier for $\psi_1$, similar to the black hole horizon. Hence, inspecting an off-shell particle configuration during the zigzag period in the $|L\rangle$ sector one might conclude  that $\psi_1$ particle should not be able to escape to the left. Instead, this is exactly what happens.

The analogy extends also to a collision of identical particles and becomes even more convincing in this case. Conventionally, a collision of two identical particles in the c.o.m. frame is described by a wave function on a half-line $\sigma<0$, or equivalently, by an odd function on the whole real axis. 
A description like this is not available in the present case.
A new pocket of space (the zigzag region) opens up at positive $\sigma$ and its size depends on the energy of colliding particles, similarly to the black hole interior.

If instead one insists on a conventional local description with odd wave functions, then one gets either the $\sigma<0$ region---a static ``black hole exterior"---with a pair of free identical fermions,  or the $\sigma>0$ region---a static ``black hole interior" with a confining potential.
In fact, this is exactly how two-particle excitations in the $\theta$-vacuum were identified in the earlier discussion presented in \cite{Paniak:1996zn}. The family of states (\ref{m}), (\ref{L}), (\ref{R}) were interpreted there as 
two families of confined mesons and a pair of unbound massive quarks.
 The fake worldsheet meson is analogous to an eternal static Schwarzschild black hole.  Both objects exist  only in the strict infinite mass limit.
 Instead, as we saw, at a finite mass there is no confined meson on the worldsheet---a zigzag appears only as a transient 
  period in the collision process.
  Importantly, it does not correspond to a resonance.
  
  Pushing this analogy  further, one might  try to identify meson quantum levels with black hole microstates.
  Then one obtains \cite{Coleman:1976uz} that there are
  \be
  {\cal N}(E)= {E^2\over \pi g^2 N}
  \ee
  states with energy smaller than $E$, which translates into
  \be
  \label{entr}
  {\cal S}=\log {E^2\over g^2 N}
  \ee
 for the entropy. Given that the size $R$ of the zigzag grows linearly with the collision energy
 \[
 R={2 E\over g^2 N}\;,
 \]  
 the expression (\ref{entr}) for the entropy nicely agrees with the logarithmic behavior of the geometric entanglement entropy in two dimensions\footnote{Note, however, that this is not a special property of a linearly growing potential. A harmonic oscillator would do equally well.}. 
 On the other hand, the corresponding  temperature is $T=E$, so that one does not expect to find Hawking radiation from a black hole with such an entropy---a single ``Hawking quantum" leads to a complete evaporation. Of course, we also did not see any sign of  copious particle production in our calculations. We ignored multiparticle processes, but these appear to be suppressed anyway similarly to the suppression of annihilations in the  identical particles case. It will be interesting to study a particle spectrum produced in the presence of light quarks with masses $m\lesssim\ell_s^{-1}$.

 It is not clear  whether the absence of Hawking radiation should be considered as a deficiency of this model, or rather a generic property of two-dimensional gravity.
 Another natural possibility in two dimensions is that the Hawking temperature is a constant set by the Planck length $\ell_s$  \cite{Fiola:1994ir}, which in our case equals to the string tension (\ref{tension}), $T\sim\ell_s^{-1}$.
 
  Interestingly, both scenarios imply the same answer for a time delay resulting from the gravitational scattering. In the worldsheet theory the time delay is proportional
  to the energy of the colliding particles (in the relativistic limit) as a shear consequence of the black hole size (the same is true for the scattering on the worldsheet of fundamental strings  \cite{Dubovsky:2012wk}), while in the $T=\ell_s^{-1}$ case as a consequence of the Hawking evaporation. This makes one wonder whether the two options may be exchanged by a choice of an observer. 
  
Note that black hole complementarity is currently a rather vague notion. In the present context we understand it as the absence of a local description covering the whole spatial slice.   
 Even in a deeply non-relativistic regime it is possible to introduce a position space wave function  only in two different patches rather  than on the whole worldsheet. Hopefully, concrete models, like the one presented here, will eventually allow to make the notion of complementarity precise.

To conclude,  the black hole information paradox \cite{Polchinski:2016hrw} is commonly viewed (and rightly so) as a conflict between locality and quantum mechanics.
The present discussion suggests that perhaps both need to be modified. In the zigzag black hole it is a violation of locality which allows quantum particles to acquire off-shell identity, leading to a version of complementarity. However, both locality and quantum mechanics are violated here in a very gentle way, so that no drama arises. 
 
\section{Comparison to  $T\bar{T}$}
\label{sec:TT}
Let us turn back to the confining string side of the story and discuss what kind of answer one may expect for the full worldsheet $S$-matrix.
Let us start with a single quark flavor. In our calculations we ignored multiparticle processes. However, the simplest expectation seems to be that, at least for heavy quarks,  high energy scattering is dominated by time delays related to zigzag processes with multiparticle production being the same as in a conventional weakly coupled quantum field theory.

An exact $S$-matrix realizing this behavior was constructed in \cite{Dubovsky:2013ira}. Namely, given an  $S$-matrix of an arbitrary relativistic two-dimensional quantum field theory  one defines a new gravitationally dressed $S$-matrix,
\be
\label{USU}
\hat{S}=USU\;,
\ee
where $U$ is a unitary operator acting on an arbitrary Fock state characterized by a set of momenta $\{k_i\}$ as
\be
\label{U}
U|\{k_i\}\rangle=e^{i\ell_s^2/4\sum_{i<j}k_i*k_{j}}|\{k_i\}\rangle
\ee
with
\[
k_i*k_j=\epsilon_{\alpha\beta}k^\alpha_ik^\beta_j
\]
and the momenta in (\ref{U}) are ordered according to their rapidities. This gravitational dressing can be recast into the operator language as a $T\bar{T}$-deformation 
\cite{Smirnov:2016lqw,Cavaglia:2016oda}, which allows to calculate a finite volume spectrum, generalizing the result for dressed free massless bosons  \cite{Dubovsky:2012wk}. Recently, the dressed $S$-matrix was derived  \cite{Dubovsky:2017cnj}  from the Jackiw--Teitelboim (JT) gravity \cite{Jackiw:1984je,Teitelboim:1983ux} coupled to an undeformed matter theory. Holographic versions of this construction were discussed in \cite{McGough:2016lol,Kraus:2018xrn,Cottrell:2018skz} and various generalizations in \cite{Smirnov:2016lqw,Giveon:2017nie,Guica:2017lia,Giribet:2017imm,Asrat:2017tzd,Cardy:2018sdv}.

Gravitationally dressed $S$-matrices exhibit many welcome features to describe the scattering on the worldsheet of confining strings in QCD$_2$ (and in higher dimensions).  Setting an undressed $S$-matrix to unity\footnote{For an asymptotically free undeformed theory this is always justified in the UV.}, and restricting to the equal mass case, the
 two-to-two scattering described by (\ref{USU}) reduces to
\be
\label{s4m}
e^{2i\delta}=e^{i\ell^2_s\sqrt{s(s-4m^2)}/4}\;.
\ee
In the ultrarelativistic limit (\ref{s4m}) turns into the shock wave phase shift $e^{i\ell_s^2s/4}$. The resulting time delay grows linearly with the collision energy. This behavior is a perfect match to (\ref{Vid}) and  a smoking gun of the string dynamics. In particular, this phase shift describes worldsheet scattering of critical strings. 
Furthermore, as an extra support for the black hole interpretation of the zigzag, this phase shift is also associated with a maximally chaotic behavior \cite{Maldacena:2015waa}. 

The finite volume spectrum associated with the gravitational dressing exhibits a characteristic square root singularity for the ground state Casimir energy\footnote{Here we assume thermal (antiperiodic) boundary conditions for fermions.} \cite{Dubovsky:2012wk,Smirnov:2016lqw,Cavaglia:2016oda},
\be
\label{EofR}
E_0(R)\sim \ell_s^{-2}\sqrt{R^2-R_c^2}\;.
\ee
This singularity can be understood as a Hagedorn phase transition related to the rapid growth of the density of states caused by the attractive interaction associated with the shock wave phase shift. Surprising at first sight, 
this singularity is however a necessary ingredient  for the worldsheet theory. It indicates that the worldsheet theory looses its meaning for compactification scales $R<R_c$ (equivalently, cannot be heated up to temperatures $T>T_c$) due to the deconfinement phase transition, associated with the spontaneous center symmetry breaking. For the adjoint QCD$_2$ the exact solution for the ground state energy in the heavy mass regime has been obtained in  
\cite{Semenoff:1996xg,Semenoff:1996ew} and it indeed takes the form (\ref{EofR}).

All these similarities make one wonder whether the worldsheet $S$-matrix can be given by a $T\bar{T}$ deformation of some local theory, perhaps a free one. In fact, the 
same question may also be asked in higher dimensions. There, however, the worldsheet dynamics is strongly restricted by a non-linearly realized 
target space Poincar\'e symmetry. In particular, a worldsheet theory at $D>2$ contains Goldstone bosons---derivatively coupled transverse coordinates $X^i$. This makes it impossible for such a simple scenario to be realized, even though there are interesting indications, coming from the lattice,  that the UV asymptotics of confining strings
at $D=3,4$  is described by 
the $T\bar{T}$ deformation  \cite{Dubovsky:2015zey}.

Restrictions related to the non-linearly realized target space Poincar\'e symmetry are absent at $D=2$. However, it appears that also at $D=2$ generically
 the worldsheet theory is not given just by the $T{\bar T}$ deformation of a local quantum field theory. To see the difference in a single fermion case let us  inspect 
 the behavior of the phase shift (\ref{s4m}) in the non-relativistic regime,
\be
2\delta\approx \ell_s^2mk\l 1+{k^2\over 2m^2}+\dots\r\;.
\ee
The corresponding time delay is 
\be
\label{dt}
\Delta t={2\d_k\delta\over v}={{\ell_s^2}m^2\over 2k}+{\ell_s^2}k+\dots\;.
\ee
The second (subleading) term in (\ref{dt}) reproduces the classical time delay corresponding to the motion in the linear potential (\ref{potential}). However, the small momentum behavior of the $T\bar{T}$ time delay is dominated by the first term in (\ref{dt}), which can be interpreted as a constant shift of a spatial coordinate. In other words a massive object in the $T\bar{T}$ story has a size which grows linearly with the mass,
\[
\Delta x={{\ell_s^2}m\over 2}\;.
\]
Such an effect does not seem to be present for heavy quarks on the worldsheet. A linear in $k^{-1}$ contribution into a time delay there,
 which gets generated both by quantum effects in the potential (\ref{potential}) and due to the annihilation contribution (\ref{Vann}), is suppressed at heavy masses.
 One may try to look for an undeformed theory, which would have a linear in $k$ contribution in the phase shift, so that the result after the dressing agrees with the worldsheet scattering. However,  a large contribution which is required,  cannot be obtained at weak coupling. Also it needs to be negative, suggesting the presence of time advances in the undeformed theory, which is at odds with locality.
 
 This discrepancy  goes away in the light mass regime, so 
 a possibility that the worldsheet theory may be described  by the $T\bar{T}$ deformation of a local theory at some special value of the mass, $m^2\sim g^2 N$, stays 
 open. In fact, there is one particularly interesting point,
 \[
 m^2={g^2 N\over\pi}\;.
 \]
 At this mass the bulk theory enjoys an accidental supersymmetry, which leaves  the trivial vacuum invariant \cite{Kutasov:1993gq}.
 However, this supersymmetry necessarily gets spontaneously broken in the $\theta$-vacuum and on the worldsheet, given that the vacuum energy (string tension) 
 is positive there. The minimal option for the worldsheet dynamics at this strongly coupled point is that, just like for heavy quark masses, there is a single particle on the worldsheet---the free quark. At this point it turns into a massless goldstino. Interestingly, a single massless fermion selfinteracting via
 the shock wave phase shift
 \be
 \label{susy}
 e^{2i\delta}=e^{i\ell_s^2s/4}
 \ee
  most likely is invariant under a non-linearly realized supersymmetry. This is similar to a non-linearly realized 
 supersymmetry for an integrable massless flow between a tricritical Ising and Ising minimal models, described by the simplest rational CDD factor\footnote{A sketch of the argument for the SUSY invariance of (\ref{susy}) proceeds as follows. This phase shift agrees with the supersymmetric one in \cite{Zamolodchikov:1991vx} at the tree and one-loop level, and all amplitudes  defined by (\ref{susy}) are soft. So one should be able to apply  arguments similar to ones presented in \cite{Dubovsky:2015zey} to show that it is supersymmetric as well. This expectation is supported by the observation that the next-to-minimal massless rational CDD factor also describes a supersymmetric RG flow \cite{Ahn:2002xp}, so it is natural to conjecture that a single fermion selfinteracting through any massless CDD factor enjoys a non-linearly realized SUSY.} \cite{Zamolodchikov:1991vx}.
 Even though there is no {\it a priori} reason for the  adjoint QCD$_2$  to be integrable at the supersymmetric point, it looks interesting to explore the possibility that it can be described just by (\ref{susy}). Note that if no new massless particles appear at this
 point, the leading order (in derivative expansion) amplitudes of the ``free" quark are completely fixed by non-linearly realized supersymmetry and integrable. So the question is mostly whether there is an obstruction to quantize QCD$_2$ consistently with this integrability.
  
 Turning to the multiflavor case one finds even more drastic differences between $T\bar{T}$-deformed theories and the worldsheet scattering. Namely, starting with an asymptotically free theory gravitational dressing results in reflectionless amplitudes at high energies. This is the opposite to what we found on the worldsheet---the two-to-two scattering there exhibits reflections and annihilations, but no transitions. The absence of transitions is a very restrictive requirement. For example, it is incompatible with integrability in the presence of a  flavor rotation symmetry \cite{Zamolodchikov:1978xm}. Note  that in principle transitions  on the worldsheet may be introduced by
deforming the gauge theory with an additional single-trace four-fermion interaction,
\[
S_{4f}=\lambda\int d^2\sigma \Tr \bar{\psi}_1\psi_2\bar{\psi_1}\psi_2\;.
\]
The possibility to add this interaction (as well as its generalizations which completely break fermion flavor symmetries, such as $ \Tr \bar{\psi}_1\psi_1\bar{\psi_1}\psi_2$) also shows that the complete orthogonality of $|L\rangle$, $|R\rangle$
states, which we found here, does not hold in general.

To conclude, it is worth noting that there is a larger class of phase shifts  exhibiting a singularity of the form (\ref{EofR}). For instance, a minimal rational bosonic CDD factor,
\be
\label{CDD}
e^{2i\delta}={\ell_s^2\sqrt{s(s-4m^2)}-2i\over \ell_s^2\sqrt{s(s-4m^2)}+2i}
\ee
also gives rise to the Hagedorn behavior \cite{Mussardo:1999aj}. The physical interpretation of these models is considered a bit of a mystery. The present setting provides a natural place to look for the physical realizations of these $S$-matrices.  For instance, (\ref{CDD}) might be a part of  an $S$-matrix 
describing  a $\theta$-vacuum of a finite $N$  integrable  two-dimensional confining gauge theory, provided such a model exists. Alternatively, (\ref{CDD}) might arise at infinite $N$ in the presence of 
quark representations which do not respect the center symmetry.

\section{Discussion}
\label{sec:last}
To conclude, the presented story has two faces---the physics of confining string for its own sake,  and its possible implications for gravity.
As far as confining strings are concerned we feel that the study of the worldsheet dynamics in QCD$_2$ is a promising approach to shed light on their puzzles.
Compared to confining strings in higher dimensions the theory appears to be less restricted and more amenable to the analytic study. This
is due to the absence of the non-linearly realized Poincare symmetry and of the corresponding Goldstone bosons---transverse modes of a string. However, much of the same physics, most notably the shock wave time delay proportional to the collision energy, arises due to the possibility of zigzags to form on the $D=2$ worldsheet.

Similar to higher dimensions, $D=2$ confining strings exhibit tantalizing similarities to the 
$T\bar{T}$ deformation, but also some differences. 
In both cases one finds a peculiar UV behavior, which was called asymptotic fragility in \cite{Dubovsky:2012wk}. The setup presented here allows
to construct many tractable examples of theories exhibiting this behavior.
For instance, it will be interesting to study a scalar adjoint QCD$_2$ with a generic single trace
potential $V=\Tr f(\phi)$. So far the general lesson seems to be that this unusual asymptotic behavior in the UV is actually caused  by semiclassical IR physics. In the gauge theory example studied here, this  is the confining potential (\ref{potential}). In the critical string case, it is the classical relation  between a proper length and a total energy for a relativistic string.

Already at this stage one can draw some preliminary lessons from QCD$_2$. Namely, in higher dimensions lattice data combined with certain theoretical considerations suggest a rather minimalistic scenario for the worldsheet theory of confining strings \cite{Dubovsky:2015zey,Dubovsky:2016cog}. We often hear a criticism that such a scenario is not realistic, and in particular contradicts to the expectations based on the experience from AdS/CFT (or better to say AdS/QCD). We do not feel that there is an actual contradiction, mainly because the present AdS/QCD analysis is done within the (super)gravity approximation, and does not tell much about strings in the actual QCD. A full-fledged string description in the background with the curvature length being the same as the string length is required.
Nevertheless it is encouraging to see that QCD$_2$ strings support the minimalistic scenario. Of course, it was already known that confining strings in two-dimensional gluodynamics are quite minimalistic---they do not carry any degrees of freedom at all. However, it is nice to see that the minimalism survives also in the presence of  dynamical worldsheet degrees of freedom.

In particular, the conventional holography suggests that there should be a strongly curved holographic three-dimensional gravitational background describing $QCD_2$. The worldsheet theory  studied here describes the worldsheet dynamics of
long fundamental strings in this background. Then one might expect to find a Liouville-like mode corresponding to the motion in the holographic direction. There is no sign of this mode neither in the adjoint QCD$_2$ nor in the pure glue theory in two-dimensions. This fits well with the absence of such a mode in $D=3,4$ lattice data. Most likely this mode disappears because the holographic description is very strongly curved similarly to how one does not find any remnant of a perturbative scalar excitation in the strongly coupled regime of the sine-Gordon model\footnote{I thank Juan Maldacena for suggesting this analogy.} \cite{Zamolodchikov:1978xm}.

Furthermore, a very interesting question, which came out in the analysis of higher dimensional confining strings, is how asymptotic freedom translates into the UV behavior of the worldsheet theory. The emerging partial answer is that the UV  worldsheet scattering is dominated by shock wave time delays. This expectation again appears to be confirmed by the adjoint  QCD$_2$ example. 

In addition, QCD$_2$ suggests a simple correspondence between the worldsheet and bulk degrees of freedom. Namely, at least in the heavy quark regime, we found that 
worldsheet excitations are in one-to-one correspondence to adjoint quarks added in the bulk theory. The operators describing the corresponding one-particle states on the worldsheet are given by
\[
{\cal O}_\psi=\Tr Pe^{i\int A}\psi\;.
\] 
Translating this expectation into $D=4$ gluodynamics, and extrapolating into the light mass regime, we expect to find three worldsheet excitations corresponding to the bulk gluon field created by 
\begin{gather}
{\cal O}_i=\Tr Pe^{i\int A}F_{zi}\nonumber\\
{\cal O}_a=\epsilon^{ij}\Tr Pe^{i\int A}F_{ij}\nonumber\;,
\end{gather}
where $z$ is a spatial direction along the string, and $i,j$ label the transverse directions. Similarly, at $D=3$ one expects to find a single excitation analogous to ${\cal O}_i$.
Interestingly, these simple expectations exactly match the Axionic String Ansatz 
of \cite{Dubovsky:2015zey,Dubovsky:2016cog}. Here ${\cal O}_i$ correspond to massless Goldstone modes, and ${\cal O}_a$ is the worldsheet axion identified in the $D=4$ lattice data in \cite{Dubovsky:2013gi}.

Coming to the gravity side, one might be skeptical on how far it will be possible to push lessons learned from such a simple lower dimensional model. We would definitely share this skepticism if presented with a rather trivial quantum mechanical system (\ref{potential}) alone as a model for black hole complementarity. It is encouraging, however,  that this model came out here from a fully relativistic theory where we expected to find gravitational physics to start with. 

Furthermore, it is well appreciated by now that the shock wave phase shift exhibits interesting gravitational features. These include the absence of local off-shell observables, related to its peculiar UV behavior \cite{Dubovsky:2012wk}, and maximal chaos \cite{Maldacena:2015waa}. The effect discussed here, which is based on a close connection between the shock wave phase shift and the physics of confinement, goes further. 
Namely, we found a global violation of locality on  large distance scales, rather than just the absence of sharply defined local observables in the UV. This is commonly expected to  be a property of quantum gravity, and it is nice to see a concrete tractable example.

Yet another
reason to be optimistic  is that even though intermediate technical steps in our analysis heavily rely on two-dimensional physics, the main outcome---complementarity as a result of off-shell identity---
is easy to formulate in any number of dimensions. Of course, it stands up as an interesting challenge now to construct a specific realization of this scenario in higher dimensions.

\section*{Acknowledgements}
I am grateful to Ofer Aharony, Nima Arkani-Hamed, Giga Gabadadze, Victor Gorbenko,   Volodya Kazakov, 
Cobi  Sonnenschein, Shimon Yankielowicz and Konstantin Zarembo for fruitful discussions. 
I thank the Weizmann Institute, where a part of this work has been done, for a warm hospitality.
This work is supported in part by the NSF CAREER award PHY-1352119.
\appendix
\section{Conventions}
\label{app:conv}
We work with the $(+,-)$ metric and choose purely imaginary $\gamma$-matrices,
\[
\gamma^0=\sigma_2=
\l\begin{matrix}
0&-i\\
i&0
\end{matrix}
\r\;,\;\;\gamma^1=i\sigma_1=\l\begin{matrix}
0&i\\
i&0
\end{matrix}
\r\;.
\]
Then the Majorana fermions are described by real Grassman fields, which have the following free mode decomposition
\[
\psi^a=\int {dk\over \sqrt{2\pi}}\l \beta_k^au(k)e^{-i\omega_k\tau +ik \sigma} +c.c.\r
\]
where
\be
\label{us}
u={i\over \sqrt{2\omega_k(\omega_k-k)}}\l
\begin{matrix}
-im\\\omega_k-k
\end{matrix}
\r
\ee
and
\[
\omega_k=\sqrt{k^2+m^2}\;.
\]
The fields in the action (\ref{action}) are linear combinations
\[
\psi=\psi^a T^a\;,\;\; F_{\mu\nu}=F_{\mu\nu}^a T^a\;,
\]
where $T^a$'s are fundamental generators of $SU(N)$ normalized 
according to 
\[
 \Tr T^aT^b={1\over 2}\delta^{ab}\;.
\]
The wave functions (\ref{us}) are normalized in such a way that 
\begin{gather}
u^\dagger u=1,\\
u_\alpha(k)u_\beta(k)^*+u^*_\alpha(-k)u_\beta(-k)=\delta_{\alpha\beta}
\end{gather}
so the canonical anticommutators are
\[
\{\beta_k^a,\beta^{b\dagger}_q\}=\delta^{ab}\delta(k-q)\;.
\]
In the two flavor case for the annihilation operators of the second flavor we use $\gamma^{a}_k$.

Occasionaly we  use a slightly non-conventional notation $\d \beta_k^{a\dagger}$, which stands for
\[
\d \beta^{a\dagger}_k|0\rangle_F=\int dk' \delta'(k-k')\beta^{a\dagger}_{k'}|0\rangle_F\;.
\]

\section{Dull Calculations with Color Indices}
\label{app:dull}
Calculations of matrix elements of the Hamiltonian (\ref{Ham}) between two-particle states repeat verbatim the ``dull calculations" in the Appendix of the Coleman's paper \cite{Coleman:1976uz} with extra color indices sprinkled here and there. 

One starts with presenting the Hamiltonian in the momentum space. The free fermion part takes the standard form\footnote{Unless specified explicitly, summations over flavor indices (equivalently, over $\beta$ and $\gamma$ oscillators) are implicit everywhere.} 
\be
H_{free}=\int d k \sqrt{k^2+m^2}\beta^{a\dagger}_k \beta^{a}_k \;,
\ee
where we subtracted an infinite vacuum energy (which corresponds to taking the $L=\infty$ limit;
otherwise we would need to keep the finite Casimir energy). The interaction with the background chromoelectric field (\ref{H2}) turns into
\be
\label{H2k}
H_2={g^2\over 2}f^{abc}(T^a-\bar{T}^a)\int dk \d \beta^{b\dagger}_k\beta^c_k+\dots\;,
\ee
where dots stand for terms changing the particle number, which do not concern us here.

To present the   gluon exchange term (\ref{H4}) in the momentum space we need the 
Fourier transform of $|\sigma-\sigma'|$,
\be
\label{absF}
|\sigma-\sigma'|=-\int {dk\over 2\pi}e^{ik(\sigma-\sigma')}\l{1\over (k-i\epsilon)^2}+{1\over(k+i\epsilon)^2} \r\equiv -2\int {dk\over 2\pi}e^{ik(\sigma-\sigma')}{{\cal P}\over k^2}\;.
\ee
Then we get
\be
H_4={g^2\over 2}\int {dk\over 2\pi}{{\cal P}\over k^2}\rho^a(k)\rho^a(-k)\;,
\ee
where 
\begin{gather}
\label{rhoa}
\rho^a(k)={-if^{abc}\over 2}\int dk'\l
\beta^b_{k'}\beta^c_{-k-k'}u_\alpha(k')u_\alpha(-k-k')
\right.\\
\left.
\nonumber
+2\beta^{\dagger b}_{k+k'}\beta^c_{k'}u^*_\alpha(k+k')u_\alpha(k')+\beta^{\dagger b}_{k'}\beta^{c\dagger}_{k-k'}u^*_\alpha(k')u^*_\alpha(k-k')
\r
\end{gather}
\subsection{Distinguishable Particles}
For the two-flavor states only the second term in (\ref{rhoa}) contributes, and the resulting Hamiltonian takes the following simple form
\be
\label{H4fk}
H_{4,2f}=g^2f^{ab_1c_1}f^{ab_2c_2}\int {dk\over 2\pi}dk_Adk_B{{\cal P}\over k^2}\beta^{b_1\dagger}_{k_A-k}\gamma^{b_2\dagger}_{k_B+k}\beta^{c_1}_{k_A}\gamma^{c_2}_{k_B}u^*_\alpha(k_A-k)u_\alpha(k_A)u^*_{\alpha'}(k_B+k)u_{\alpha'}(k_B)
\ee

Using (\ref{H2k}) and (\ref{H4fk}) it is straightforward to check that both $H_2$ and $H_4$ are diagonal
in the $|L\rangle$, $|R\rangle$ basis.
Their non-vanishing entries are 
\begin{gather}\label{H2fin}
\langle k_1',k_2',L|H_2|k_1,k_2,L\rangle=-\langle k_1',k_2',R|H_2|k_1,k_2,R\rangle=-i{g^2N\over 4}\delta'(k_1-k_1')\delta\l\sum k\r\\
\label{H4fin}
\langle k_1',k_2',L|H_4|k_1,k_2,L\rangle=\langle k_1',k_2',R|H_4|k_1,k_2,R\rangle=
-{g^2N\over 4\pi}{{\cal P}\over (k_1-k_1')^2}{\cal U}\delta\l\sum k\r
\end{gather}
where
\[
\sum k=k_1+k_2-k_1'-k_2'
\]
and
\be
\label{calU}
{\cal U}=u^*_\alpha(k'_1)u_\alpha(k_1)u^*_{\alpha'}(k'_2)u_{\alpha'}(k_2)={(k_1k'_1+\omega_{k_1}\omega_{k'_1}+m^2)^{1/2}(k_2k'_2+\omega_{k_2}\omega_{k'_2}+m^2)^{1/2}\over 2(\omega_{k_1}\omega_{k_2}\omega_{k'_1}\omega_{k'_2})^{1/2}}\;.
\ee
Note that ${\cal U}=1$ both in a non-relativistic regime, when all momenta $k\ll m$, as well as in the ultra-relativistic one, $k\gg m$.
\subsection{Identical Particles}
For identical particles the $H_2$ Hamiltonian acts  on a two-particle state as
\begin{gather}\label{H2in}
H_2|k_1,k_2\rangle=-i{g^2N\over 4}{2\over N^{3/2}}\l \d\beta^{a_1\dagger}_{k_1}\beta_{k_2}^{a_2\dagger}-
\beta^{a_1\dagger}_{k_1}\d\beta_{k_2}^{a_2\dagger}  \r|0\rangle\otimes T^{a_1}T^{a_2}\;.
\end{gather}
This corresponds to the following matrix element
\be
\label{H2id}
\langle k_1',k_2'|H_2|k_1,k_2\rangle=-i{g^2N\over 4}\delta'(k_1-k_1')\delta\l\sum k\r\;,
\ee
which is not different from nom the two-flavor case. However, the gluon exchange interaction is more complicated now, due to a possibility of annihilations.
All terms in (\ref{rhoa}) contribute now and the resulting Hamiltonian turns into a sum of the annihilation and transition terms.
The annihilation part is
\begin{gather}
H_{a}={g^2\over 4}f^{ab_1c_1}f^{ab_2c_2}\int {dk\over 2\pi}dk_Adk_B{{\cal P}\over k^2}
\nonumber
\beta_{k-k_A}^{c_1\dagger}\beta_{k_A}^{b_1\dagger}\beta_{k_B}^{b_2}\beta_{k-k_B}^{c_2}u^*_\alpha(k_A)u^*_\alpha(k-k_A)u_\beta(k_B)u_\beta(k-k_B)
\end{gather}
and the transition one is
\be
\nonumber
H_t={g^2\over 2}f^{ab_1c_1}f^{ab_2c_2}\int {dk\over 2\pi}dk_Adk_B{{\cal P}\over k^2}
\beta_{k_A-k}^{b_1\dagger}\beta_{k+k_B}^{b_2\dagger}\beta_{k_A}^{c_1}\beta_{k_B}^{c_2}u^*_\alpha(k_A-k)u_\alpha(k_A)u^*_\beta(k+k_B)u_\beta(k_B)\;.
\ee
The annihilation matrix elements are
\be
\label{H4an}
\langle k_1',k_2'|H_a|k_1,k_2\rangle={g^2N\over 4\pi}{{\cal P}\over (k_1+k_2)^2}{\cal U}_{a}\delta\l\sum k\r
\ee
where
\[
{\cal U}_a={(k_1k_2+\omega_{k_1}\omega_{k_2}-m^2)^{1/2}(k'_1k'_2+\omega_{k_2}\omega_{k'_2}-m^2)^{1/2}\over 2(\omega_{k_1}\omega_{k_2}\omega_{k'_1}\omega_{k'_2})^{1/2}}\;.
\]
Note that there is no $s$-channel pole in (\ref{H4an}) because ${\cal U}_a$ vanishes at $k_1+k_2\to 0$ and cancels the singularity in (\ref{H4an}),
producing a non-vanishing finite result for the matrix element at $k_1+k_2=0$.

Finally, for the transition matrix elements we obtain
\be
\label{H4tr}
\langle k_1',k_2'|H_t|k_1,k_2\rangle=-{g^2N\over 4\pi}
{{\cal P}\over (k_1-k'_1)^2}
{\cal U}\delta\l\sum k\r\;.
\ee
where ${\cal U}$ is given by (\ref{calU}).

\bibliographystyle{utphys}
\bibliography{dlrrefs}
\end{document}